# Generation of THz radiation through molecular modulation in hydrogen-filled hybrid anti-resonant fibers

SÉBASTIEN LORANGER,[1,*] FOROOGH JAFARI,[1] JOSEBA ZUBIA,[2,3] AND DAVID NOVOA,[2,3,4]

[1]Electrical Engineering department, Polytechnique Montreal, 2900 Edouard-Montepetit, Montreal H3T 1J4, Canada
[2]Department of Communications Engineering, Engineering School of Bilbao, University of the Basque Country (UPV/EHU), Torres Quevedo 1, 48013 Bilbao, Spain
[3]EHU Quantum Center, University of the Basque Country (UPV/EHU), 48013 Bilbao, Spain
[4]IKERBASQUE, Basque Foundation for Science, Plaza Euskadi 5, 48009 Bilbao, Spain
*sebastien.loranger@polymtl.ca

**Abstract:** We study the generation of narrowband terahertz (THz) pulses by stimulated Raman scattering and molecular modulation in hydrogen-filled hybrid hollow-core fibers. Using a judicious combination of materials and transverse structures, this waveguide design enables simultaneous confinement of optical and THz signals with reasonably low attenuation, as well as high nonlinear overlap. The THz pulses are then generated as the second Stokes band of a ns-long near-infrared pump pulse, aided by Raman coherence waves excited in the gaseous core by the beat-note created by the pump and its first Stokes band. Optimization of the fiber characteristics facilitates phase matching between the corresponding transitions and coherence waves while avoiding coherent gain suppression, resulting in optical-to-THz conversion efficiencies up to 60%, as confirmed by rigorous numerical modelling under ideal conditions. When the current optical material constraints are considered, however, the attainable efficiencies relax to 0.2%, a still competitive value compared to other systems. The approach is in principle power and energy scalable, as well as tunable in the 1 – 10 THz range without any spectral gaps, thereby opening new pathways to the development of fiber-based THz sources complementary to other mature technologies such as quantum cascade lasers.

## 1. Introduction

Terahertz (THz) radiation has many applications in, for example, characterization of materials, sensing and wireless communications with significant potential in the next generation 6G and beyond [1]. Despite their widespread interest, progress in those fields has been hampered by the limited number of tools available for the generation and manipulation of THz waves. Among other solutions, broadband THz pulses can be generated via optical rectification or difference-frequency generation of near-infrared (NIR) ultrashort pulses [2-4]. These broadband THz pulses have been successfully employed to study materials' properties [5] and condensed matter physics [6], but are unfortunately not suitable for all applications. For instance, in telecoms [7] or metrology [8], a narrowband emission may be more adequate. In this regard, gyrotrons were recently proposed as continuous-wave or pulsed narrowband THz sources [9]. They are, however, bulky and complex systems, which may limit their ultimate widespread deployment. Quantum cascade lasers (QCLs) [10] are currently the most flexible sources of narrowband radiation in the 1–3 THz range with reasonable efficiency and brightness. These can deliver powers in the mW-level but require cryogenic temperatures to operate, which makes them cumbersome and not very cost-effective. To partially solve this issue, sub-mW room-temperature QCLs in the THz regime have already been demonstrated through difference-frequency generation [11, 12].

In this work we propose a novel route to generate few-THz signals through stimulated Raman scattering (SRS) and molecular modulation in hydrogen-filled hybrid anti-resonant fibers pumped in the NIR [13]. Among all the Raman-active molecules in nature, $H_2$ displays the highest Raman frequency shift ($\Omega \sim 125$ THz) and gain owing to the strong covalent bond and lightweight of its atomic constituents [14-17]. Such unique properties can then be exploited to down-convert pump light in the optical domain to a few THz. The direct first-Stokes conversion is, however, extremely inefficient as the effective gain is proportional to the generated frequency, which in the case of THz is well below the quantum defect. Interestingly, such weak SRS-driven interaction can be boosted by the use of coherent vibrational excitations pre-existent in the gaseous core [18, 19]. This technique, known as molecular modulation, has been proven useful in the generation of multi-octave-spanning Raman combs [20], as well as efficient frequency conversion of broadband signals [21], ultraviolet light [22] and even single photons [23]. Efficient molecular modulation relies on the possibility of arranging phase matching between the excited optical phonons or coherence waves and the corresponding optical transitions of interest, as well as their effective nonlinear overlap. In this work we propose a hybrid hollow-core anti-resonant fiber (HARF) design capable of confining both optical and THz modes so that they significantly overlap and collinear phase matching between the interacting fields is possible. Using this approach, we found that the theoretical quantum efficiency (QE) of the THz generation process can exceed 60% in the absence of waveguide losses. Extending the analysis to a more realistic scenario incorporating the lowest-loss THz-compatible materials available to date, the QE drops to 0.2%, a reasonably high value only limited by the current technology. In addition to the high QEs attainable, this frequency down-conversion nonlinear process is scalable in energy, power (through the use of high-repetition rate laser systems) and frequency by using a tunable pump laser, thereby opening a new route to the efficient generation of multi-mW narrowband THz pulses, complementary to the existent QCL technology.

## 2. Theory and simulations

When an optical narrowband pump pulse propagates in a Raman-active molecular medium, it can be inelastically scattered off giving rise to a Stokes beam down-shifted by the Raman transition frequency $\Omega$. This "self-phase-matched" process is accompanied by the excitation of a coherence wave of synchronous molecular motion that triggers further amplification of the Stokes band via SRS. Such collective molecular excitation can then be used to stimulate frequency conversion between any other pair of frequencies offset by $\Omega$, provided certain phase-matching conditions are met. On the other hand, when attempting to use direct SRS in $H_2$ to down-shift optical radiation to THz frequencies, the Raman gain is so low that the required pump beam in the mid-infrared (MIR) is simply incapable of generating from noise on its own strong enough molecular coherence at reasonable pump power levels. However, it is possible to circumvent this issue using molecular modulation assisted by another coherence wave generated by the beat-note triggered by an NIR pump pulse and its first-Stokes band in MIR. In this scenario, the resulting optical phonons can then be used to convert photons from the MIR Stokes band to a second Stokes lying in the THz domain without threshold, as shown in Fig. 1. For this to happen, the dephasing $\Delta\beta = (\beta_P - \beta_{S1}) - (\beta_{S1} - \beta_{S2}) = \beta_p + \beta_{S2} - 2\beta_{S1}$, where $\beta_p, \beta_{S1}, \beta_{S2}$ are the propagation constants of the NIR pump (P), first MIR Stokes (S1) and second THz Stokes (S2), respectively, must be made vanishingly small. Importantly, in order to be efficient this process will have to compete with the up-conversion by $\Omega$ of the pump to its first (AS1) and second (AS2) anti-Stokes bands. This can be realized by designing the hybrid waveguide so as to ensure large dephasing between any anti-Stokes-related transitions and the coherence waves.

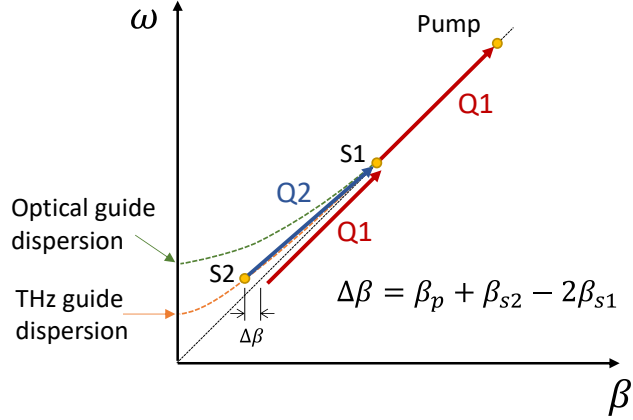

Fig. 1: Sketched dispersion diagram (not to scale and exaggerated dispersion) showing the different signals and coherence waves (S1: 1st Stokes, S2: 2nd Stokes, Q: coherence wave) interacting with a slight non-zero dephasing. The contributions of the anti-Stokes transitions to the overall coherence are not shown but included in the calculation as a negligible oscillating component of Q1. Under perfect phase matching ($\Delta\beta = 0$), Q1 and Q2 have the same linear momentum but different overlap factors.

To validate the viability of our approach, we simulated the nonlinear dynamics of the fields propagating inside a hydrogen-filled HARF (its specific design will be discussed in the next sections) using the following set of coupled Maxwell-Bloch equations [21]:

$$\frac{\partial E_l}{\partial z} = -i\kappa_{2,l}\frac{\omega_l}{\omega_{l-1}}E_{l-1}q_{l-1}q_l^* \sum_M \frac{s_{l,m}}{s_{Q,m}}Q_m - i\kappa_{2,l+1}E_{l+1}q_{l+1}q_l^* \sum_M \frac{s_{l+1,m}}{s_{Q,m}}Q_m^* - \frac{1}{2}\alpha_l E_l, \quad (1)$$

$$\frac{\partial Q_m}{\partial \tau} + \frac{Q_m}{T_2} = -i\frac{1}{4}s_{Q,m}\kappa_{1,m}E_m E_{m-1}^* q_m q_{m-1}^*, \quad (2)$$

where $l$ is an integer denoting the pump ($l = 0$) and its sidebands with angular frequencies $\omega_l = \omega_P + 2\pi\Omega l$, where $\omega_P$ is the pump frequency and $-2 < l < 2$ in the case studied here. Fields with indices outside of this range are not considered (e.g. $E_{-3} = E_3 = 0$). The complex field is described as $e(x,y,z,\tau) = F_l(x,y)E_l(z,\tau)q_l$ with $F_l$ as its normalized transverse profile, $E_l$ the slowly varying electric field envelope and $q_l = \exp(-i\beta_l z)$ the fast-oscillating spatial phase term involving the modal propagation constants $\beta_l$. $\alpha_l$ represents the propagation loss, $Q$ is the amplitude of the Raman coherence waves and $\tau$ is the time. The coupling constants are $\kappa_{1,l} = c\varepsilon_0[2g_l/(NT_2\hbar\omega_{l-1})]^{0.5}$ and $\kappa_{2,l} = \kappa_{1,l} N\hbar\omega_{l-1}/(2c\varepsilon_0)$ with gain values $g_l$, molecular number density $N$ and coherence dephasing time $T_2$ [14]. The nonlinear spatial overlaps $s_{Q,m}$ defining the coherence wave transverse profile can be absorbed by redefining an effective $Q_m^{eff} = Q_m/s_{Q,m}$. The overall nonlinear spatial overlap integrals accounting for the strength of the nonlinear coupling between the interacting fields and coherence waves are:

$$s_{l,m} = \begin{cases} 1 & m = 0,1,2 \quad l = 0,1,2 \\ s_1 = \frac{\int F_{opt}^3 F_{THz} dA}{\int |F_{opt}|^2 dA \left(\int |F_{opt}|^2 dA \int |F_{THz}|^2 dA\right)^{0.5}} A_{eff} & m = 0,1,2 \quad l = -1 \\ s_2 = \frac{\int F_{opt}^2 F_{THz}^2 dA}{\int |F_{opt}|^2 dA \int |F_{THz}|^2 dA} A_{eff} & m = -1 \quad l = -1 \end{cases} \quad (3)$$

Where $F_{opt}$ and $F_{THz}$ are the real-valued normalized transverse optical and THz fields, respectively, and $A_{eff}$ the effective mode area of the pump wave. The first two possibilities of $s_{l,m}$ represent the overlap with the same coherence wave, i.e. the optically generated phonons using the same $s_{Q,m}$ term, and can therefore be grouped into a single coherence wave Q1. On the other hand, the last possible value of $s_{l,m}$ represents the overlap of the optical pump field with the THz-generated coherence wave, which will have a different transverse profile. Because of this difference, the related $Q_{-1}^{eff}$ must be considered as a second independent coherence wave Q2. Note that this last contribution is extremely small, but does play a role in low-loss, long propagation scenarios. Nonlinear overlap among the interacting optical waves is assumed to be perfect since they propagate in frequency-independent LP$_{01}$-like modes limited by the physical wall diameter.

Fig. 1 depicts a schematic (not to scale) dispersion diagram of the electromagnetic and coherence waves considered in the calculation. As a generalization, a small phase-mismatch is considered in this figure to distinguish between Q1 and Q2. The role of this mismatch will be discussed in the next section.

### 3. THz generation in gas-filled hybrid anti-resonant fiber: Theoretical limit and coherent gain suppression

The design of the hollow waveguide is critical to allow efficient THz generation via SRS and molecular modulation. This is because three hardly compatible conditions must be met: low loss, high nonlinear overlap and phase matching among the interacting fields and coherence waves. Unfortunately, simultaneous fulfilment of these conditions is nowadays difficult since, in the current state of the art, there is no THz-transparent material capable of providing low-loss (< 1 dB/m) optical and THz transmission. We strongly believe, nevertheless, that this is not an actual fundamental limit of this approach, but rather a current technological limitation. Therefore, we propose a HARF concept where the central hollow region contains a microstructured cladding capable of confining the NIR pump and MIR Stokes modes via anti-resonant reflection [24], whereas a strong-confining structure such as a photonic bandgap pattern, Bragg-reflecting rings and/or metallic layer, keeps the THz wave inside the hollow region on top of the anti-resonant fine-structure, as shown in Fig. 2.

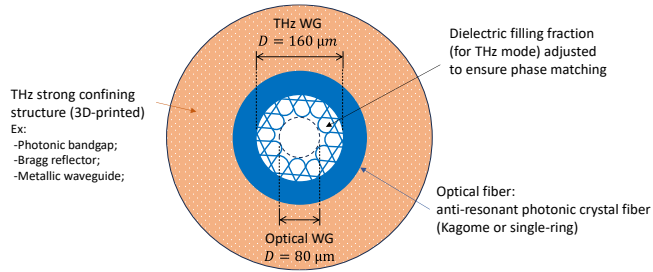

Fig. 2: Sketch of transverse structure of the proposed ideal HARF. The hollow optical fiber is encapsulated in a mm-size structure (that can be 3D printed) which acts as a strongly confining waveguide (WG) for the THz waves. The optical field is guided by an anti-resonant fine-structure in the hollow region (kagome or single-ring).

In this theoretical study, we will analyze the impact of waveguide losses on the performance of fiber-based THz generation by considering THz losses in the $0 - 20$ dB/m range. Attenuation of the optical fields confined in the hollow channel of the HARF will be totally disregarded, since loss values well below 0.1 dB/m have already been demonstrated both for hypocycloid-contour hollow-core kagome photonic crystal fibers [25] (as that employed in our study, see Fig. 2) or simplified single-ring anti-resonant structures [26].

As for the nonlinear overlap integrals between the optical and THz waves (Eq. 3), it is not straightforward to establish a theoretical limit, as one can always try to conceive a better or more optimized waveguide. In our case, high overlap requires the optical fields to be loosely confined (optical hollow-core diameter >> wavelength) and the THz field to be tightly confined (external waveguide diameter close to wavelength). The latter condition requires operation close to the cut-off for the THz modes, which forces their effective index to drop making it necessary to significantly overlap with some high-index material to minimize dephasing. With these constraints in mind, our realistic compromise solution is a HARF composed of a THz cylindrical waveguide ($TE_{11}$ mode) using a Bragg-reflecting or photonic-bandgap structure of diameter twice that of the optical hollow-core waveguide ($LP_{01}$-like mode), provided a significant low-loss material is available. In such a design the nonlinear overlap integrals are $s_1 = 0.58$ and $s_2 = 0.35$, calculated using the optical and THz fields obtained via finite-element modelling (FEM).

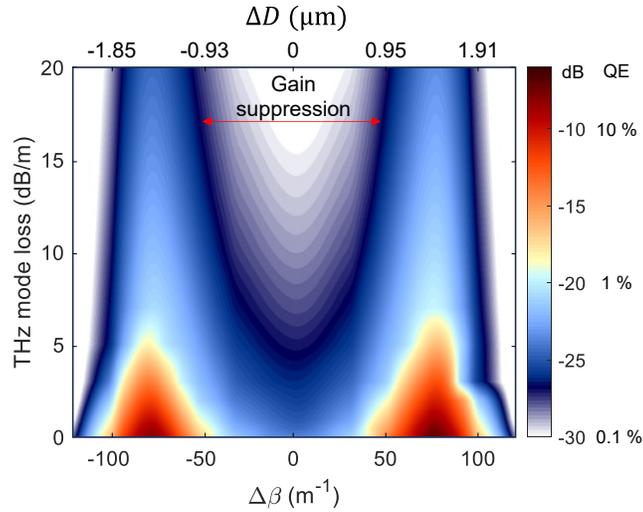

Fig. 3: Simulated quantum efficiency map for an ideal 3-m long HARF filled with 70 bar $H_2$ and pumped with 3 ns, 0.5 mJ pulses. Only losses of the THz mode are considered. The maximum QE in a lossless fiber is ~20% for a dephasing of 75 $m^{-1}$. Note that the QE drops dramatically in the vicinity of the perfect phase-matching points owing to the onset of coherent gain suppression. The horizontal axis on top shows to which values of THz waveguide diameter $D$, represented by the variation $\Delta D = D - D_0$, where $D_0 = 160$ μm, would correspond the phase mismatch $\Delta\beta$.

Using the model represented by Eqs. (1–3), we evaluated the performance of our ideal system to generate a narrowband pulsed second Stokes band at 3 THz (S2), obtained by pumping the $H_2$-filled HARF described above with 3 ns pulses of 1189 nm central wavelength (an easily accessible wavelength in commercial optical parametric oscillators) and generating a first MIR Stokes at 2350 nm (S1). Note that the system is fully tunable in the 1 – 10 THz range by slightly adjusting the pump wavelength in the 1157 – 1199 nm range. Without loss of generality, the $H_2$ filling pressure is fixed at 70 bar in all simulations to ensure saturation of the material Raman gain [14]. It is assumed that the hybrid waveguide can always be engineered to be near the desired $\Delta\beta$ condition and then fine-tuned through gas pressure.

The results of a series of simulations of the nonlinear propagation of 0.5 mJ pump pulses along a 3 m-long HARF filled with 70 bar $H_2$ is shown in Fig. 3. The false color map displays optical-to-THz QE as a function of both THz mode losses and dephasing. Interestingly, at perfect phase matching ($\Delta\beta = 0$) and when THz losses are disregarded (i.e., an *a priori* ideal nonlinear conversion scenario), the maximum QE is ~ 0.45% and drops even further as losses start to raise. This counterintuitive phenomenon is called coherent gain suppression [27, 28] and happens when the rates of phonon creation in pump-to-Stokes transitions and phonon annihilation in pump-to-anti-Stokes transitions precisely balance. In our system, this occurs when the amplified S1 acts as pump for S2, in which case the new Stokes generation S2 is phase-matched with anti-Stokes generation towards the original pump. However, since the original pump is almost fully depleted after S1 generation while S2 is still very weak (see Fig. 4), a balance between the pump and S2 is reached, inhibiting further growth of Raman coherence. To mitigate this effect, we simply impose a small dephasing, easily achievable by slightly adjusting the structural design. With this modification, the QE largely increases to a surprising value ~ 20% for $\Delta\beta \sim 75$ m$^{-1}$, not yet limited as the THz signal is still experiencing gain after 3 m, as it can be seen in the propagation dynamics displayed in Fig. 4. In this ideal scenario, we have verified that above 60% conversion could be reached with sufficient HARF length (see Fig. 4e), which would be nevertheless unrealistic under the current technological limitations discussed before. Remarkably, it turns out that such high QE values are still attainable using other strategies such as increasing pump pulse energies, as will be shown later.

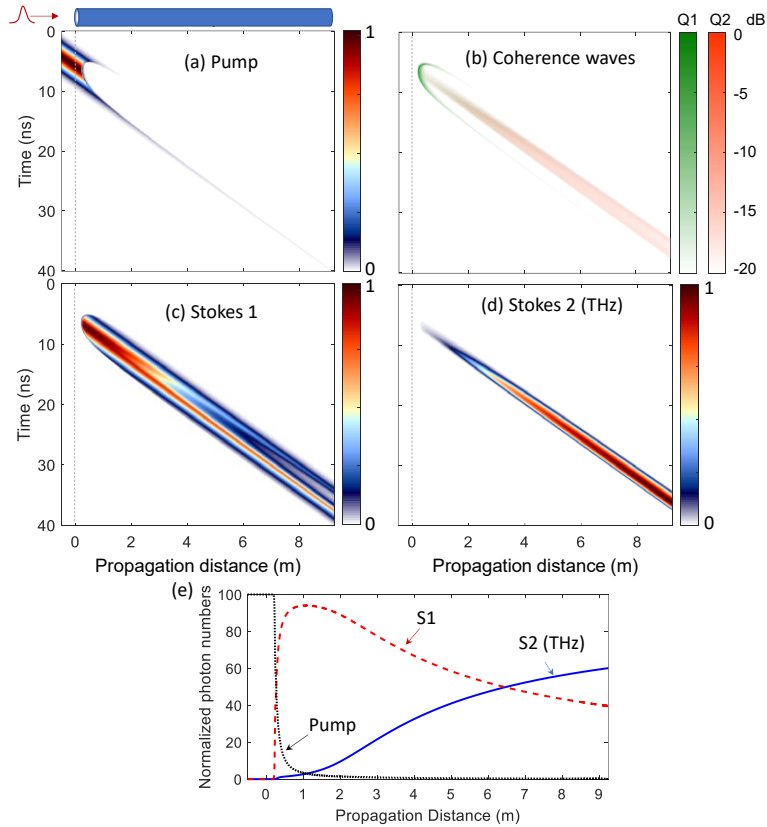

Fig. 4: Simulated nonlinear Raman conversion dynamics in an ideal 3-m long HARF filled with 70 bar H$_2$ and pumped with 3 ns, 0.5 mJ pulses. The pump pulse is initially located outside the fiber, whose input end lies at 0 m. The different panels show the normalized spatio-temporal evolution of (a) the NIR pump pulse, (b) the Raman coherence waves, (c) the first MIR Stokes signal (S1) and (d) the second THz Stokes band (S2). (e) Quantum efficiency of the optical-to-THz conversion process (represented as normalized photon numbers in percentage) as a function of the position along the HARF.

Further insight into the in-fiber nonlinear frequency conversion process can be gained by analyzing the complex pulse evolution dynamics displayed in Fig. 4. In fact, we can clearly identify two distinct regimes of frequency conversion, governed by both Q1 and Q2 coherence waves, respectively. In the first regime, a short and intense coherence wave Q1 stimulates the growth of a coherent seed of the S2 THz band. This first process occurs over a very short fiber length and is characterized by a nearly complete energy transfer from the pump to S1. Due to the quick depletion of the pump, inevitable under high-energy pumping, the coherence wave Q1 is very short-lived, thus limiting the growth of S2 and making this first stage rather loss-independent. The second regime corresponds to the further amplification of the coherent S2 signal by SRS using the strong S1 signal as a pump. As discussed above, this low-gain process is strongly dependent on the dephasing since small $\Delta\beta$ will result in the onset of coherent gain suppression. In contrast to the "short" first regime, this subsequent amplification stage relies on long interaction lengths to compensate for the low gain inherent to operation in the vicinity of the quantum defect, hence being very loss-dependent.

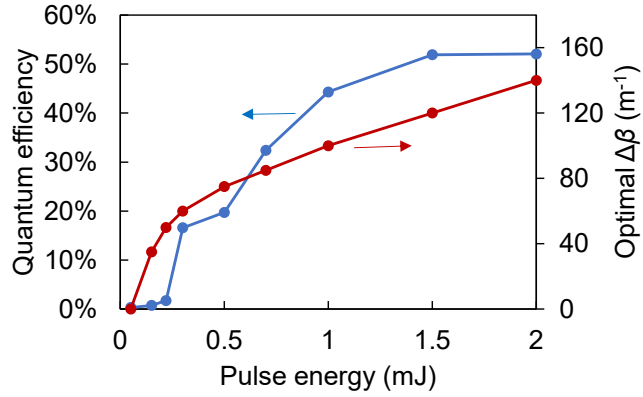

Fig. 5: Maximum quantum efficiency (blue solid line) and corresponding optimal dephasing (red solid line) for the loss-less system shown in Fig. 2 and explained in the text, as a function of pump pulse energy.

As this is a highly nonlinear process, strong pump pulses are required to achieve reasonable conversion efficiency. As mentioned above, we consider a commercially-available tunable Q-switched laser source capable of emitting 3 ns-long NIR pulses with energies up to 2 mJ. In Fig. 5 we display the maximum optical-to-THz QE obtained for the system described above as a function of input pump pulse energy. As expected for a nonlinear system, the efficiency changes with pump energy, reaching QE up to ∼ 50 % at 1.5 mJ. Along with the QE, we plot the optimal dephasing needed to achieve it, which will in general depend on the pump pulse characteristics. Note that for increasing pump energies, a larger $\Delta\beta$ is required to attain maximum QE, in excellent agreement with the theory of coherent gain suppression [28].

To assess the performance of the HARF-based optical-to-THz convertor, it is important to analyze not only the QE but the overall THz output power attainable, i.e. the power efficiency (PE) of the system given by:

$$PE(\%) = \frac{f_{THz}}{f_{Pump}} QE(\%). \quad (4)$$

This expression indicates how PE is intrinsically limited by the quantum defect, which is very detrimental in SRS-based THz generation. In our system, the PE for the generation of 3 THz pulses is proportional to ∼ 0.012 QE (%), meaning that a 60% QE would yield a ∼ 14 mW THz output for a standard pump average power of 2 W (2 mJ pulses at 1 kHz repetition rate).

In general, as displayed in Fig. 5, the higher the input pump energy, the higher the conversion efficiency. In a real scenario, however, such scaling might be limited by the onset of thermal instabilities as well as detrimental nonlinear effects – which might also appear if shorter pulse durations are employed.

## 4. Realistic propagation conditions under current technological constraints

In the previous section we discussed how a novel hydrogen-filled HARF may be used to efficiently generate and confine narrowband THz signals. The ideal nonlinear system was optimized to simultaneously fulfill, to a large extent, the three conditions presented above: low loss, high nonlinear overlap and phase matching. Such design, however, is challenging to realize under the state-of-the-art, mainly because of the lack of low-loss, fiber-compatible THz-transparent materials. In this section we will analyze a realistic version of the fiber structure displayed in Fig. 2 considering the current technological limitations (see Fig. 6a). In this regard, two of the lowest loss, fiber-compatible polymeric materials are cyclo olefin (known as Zeonor/Zeonex) and Polymethylpentene (TPX) [29, 30]. Microstructured fibers have already been fabricated using these materials [31] and therefore will be considered for our feasibility analysis. As discussed previously, the optical fields, that is the anti-Stokes, the pump (P) and the first Stokes (S1) bands are guided in the inner anti-resonant-reflecting microstructure. The wavelength-independent, fundamental modal fields obtained by FEM are shown in Fig. 6b. For the analysis of the phase-matching conditions, their dispersion can be well approximated using simplified analytical models [32].

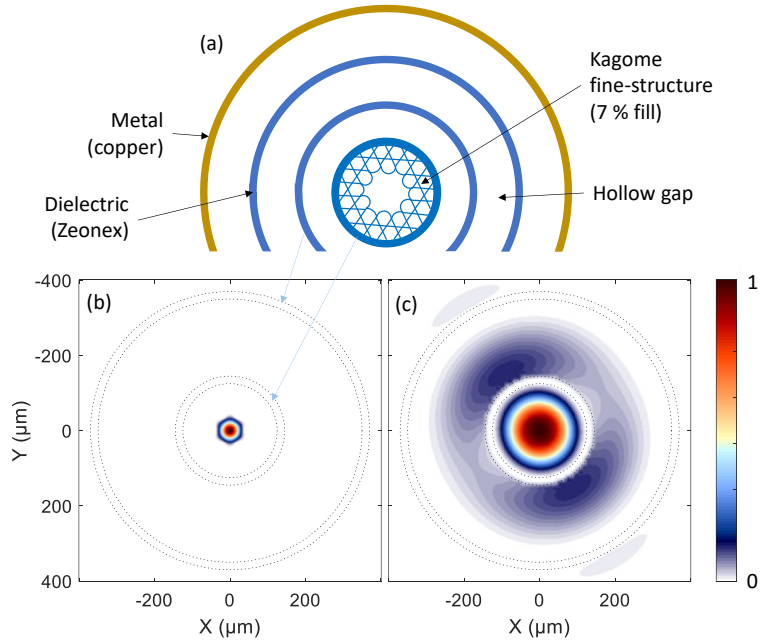

Fig. 6: a) Hybrid waveguide design (not to scale and the sketch is incomplete) composed of an anti-resonant optical waveguide and a Bragg-enhanced cylindrical metallic THz waveguide. Thickness of dielectric material and hollow gap are optimized to maximize overlap with reasonable losses. b) Simulated fundamental modal optical field compared to c) the calculated near-cut-off THz mode within the 1$^{st}$ Bragg ring. The field is negligible beyond the scale displayed. While the optical mode exhibits negligible attenuation, the THz mode has –13 dB/m loss, mainly caused by overlap with the lossy dielectric material.

The THz field, on the other hand, is mainly confined within the dense cladding of the hollow-core anti-resonant fiber, while overlapping with its thin glass microstructure (see Fig. 6c). This fine structure is approximated as a low-density homogeneous material (air + silica) in the FEM calculations. Note that everything beyond the inner, dense-cladding boundary is open to design with the goal of maximizing confinement (e.g., improving overlap) while minimizing losses. In this analysis we propose a 3D-printed Bragg waveguide structure (fine struts placed at long intervals to provide structural support are omitted for clarity) in which the optical hollow-core microstructured fiber is inserted (see Fig. 6a), thus forming a HARF. Each concentric ring is made of Zeonex and the outermost layer is a copper capillary to ensure a final high reflection. The thicknesses of both fiber cladding and dielectric layers are selected to fulfill anti-resonant-reflecting conditions to minimize material overlap. Numerical modelling indicated that such a design can yield losses as low as 5 dB/m, but with very poor optical-to-THz overlap.

Considering that it is difficult to achieve both regimes of Raman frequency conversion given the current material constraints, we focus on the first regime –for which losses are not critical– where a short-lived strong coherence wave is excited in the gaseous core and stimulates the growth of a coherent THz band. The tested design has an optical core diameter of 80 μm, a fine-structure dielectric filling fraction of 7%, a THz core of 250 μm diameter, air gaps of 205 μm and polymer thickness of 23 μm. The THz losses are as low as –13 dB/m (optical losses are vanishingly small), with nonlinear overlap parameters $s_1 = 0.28$ and $s_2 = 0.08$. Note that while only the first two air gaps are required to obtain the near-cut-off THz mode-field displayed in Fig. 6c, inclusion of additional Bragg-reflecting layers help to further reduce attenuation.

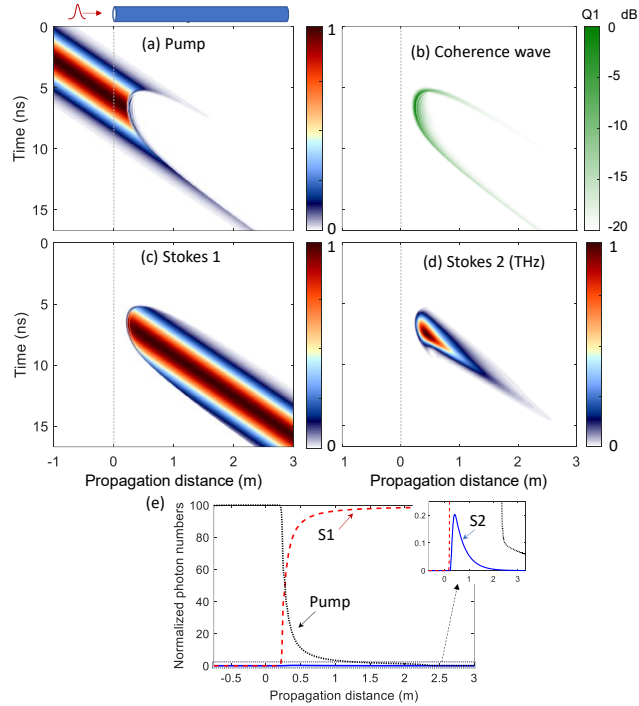

Fig. 7: Simulated nonlinear Raman conversion dynamics in a realistic 3-m long HARF filled with 70 bar $H_2$ and pumped with 3 ns, 0.5 mJ pulses. The pump pulse is initially located outside the fiber, whose input end lies at 0 m. The different panels show the normalized spatio-temporal evolution of (a) the NIR pump pulse, (b) the Raman coherence waves, (c) the first MIR Stokes signal (S1) and (d) the second THz Stokes band (S2). (e) Quantum efficiency of the optical-to-THz conversion process (represented as normalized photon numbers in percentage) as a function of the position along the HARF.

We simulated SRS-based THz generation in this realistic HARF, obtaining a maximum QE of 0.2% when the system was pumped by 3 ns-long, 0.5 mJ pulses. The field propagation results are shown in Fig. 7. It is important to highlight that, due of the moderate THz loss, the THz field is generated close to the peak of the coherence wave, reaching its maximum strength at

0.4 m. This corresponds to the optimal fiber length, as further propagation will only result in attenuation of the THz signal.

## 5. Conclusions

We proposed the use of stimulated Raman scattering and molecular modulation in hydrogen-filled hybrid anti-resonant fibers to generate narrowband pulsed THz radiation. In addition to power and energy scalability, and large gapless tunability in the 1 – 10 THz range by simple adjustment of the pump wavelength, our scheme has the potential to reach optical-to-THz quantum efficiencies above 60%. This approach is radically different, yet complementary to those traditionally used by the THz community such as e.g. continuous-wave QCLs and ultrafast optical rectification, and can therefore pave the way to the development of a new family of high-performance fiber-based THz sources.

**Acknowledgments:**

The work of JZ and DN was supported by the grants PID2021-123131NA-I00 and TED2021-129959B-C21, funded by MCIN/AEI/10.13039/501100011033, by "ERDF a way of making Europe" and by the "European Union NextGenerationEU/PRTR", and the Gobierno Vasco/Eusko Jaularitza (IT1452-22) and ELKARTEK (KK-2021/00082 and KK-2021/00092). D. N. also acknowledges support from the IKUR Strategy of the Department of Education of the Basque Government.

**References**

1. Chaccour, C., et al., *Seven Defining Features of Terahertz (THz) Wireless Systems: A Fellowship of Communication and Sensing.* IEEE Communications Surveys & Tutorials, 2022. **24**(2): p. 967-993.
2. Fülöp, J.A., S. Tzortzakis, and T. Kampfrath, *Laser-Driven Strong-Field Terahertz Sources.* Advanced Optical Materials, 2020. **8**(3): p. 1900681.
3. Yu, N.E., *Terahertz generation in quasi-phase-matching structures.* Journal of the Korean Physical Society, 2022. **81**: p. 580-586.
4. Yadav, S., et al., *Nonlinear optical single crystals for terahertz generation and detection.* Journal of Nonlinear Optical Physics & Materials, 2022. **31**(02): p. 2230001.
5. Globus, T.R., et al., *THz-Spectroscopy of Biological Molecules.* Journal of Biological Physics, 2003. **29**(2): p. 89-100.
6. Ménard, J.M., et al., *Revealing the dark side of a bright exciton–polariton condensate.* Nature Communications, 2014. **5**(1): p. 4648.
7. Abramov, P.I., et al., *Quantum-Cascade Lasers in Atmospheric Optical Communication Lines: Challenges and Prospects (Review).* Journal of Applied Spectroscopy, 2020. **87**(4): p. 579-600.
8. Consolino, L., et al., *QCL-based frequency metrology from the mid-infrared to the THz range: a review.* 2019. **8**(2): p. 181-204.
9. Idehara, T., et al., *The Gyrotrons as Promising Radiation Sources for THz Sensing and Imaging.* Applied Sciences, 2020. **10**(3).
10. Sengupta, K., T. Nagatsuma, and D.M. Mittleman, *Terahertz integrated electronic and hybrid electronic–photonic systems.* Nature Electronics, 2018. **1**(12): p. 622-635.
11. Mikhail, A.B. and C. Federico, *New frontiers in quantum cascade lasers: high performance room temperature terahertz sources.* Physica Scripta, 2015. **90**(11): p. 118002.
12. Fujita, K., et al., *Recent progress in terahertz difference-frequency quantum cascade laser sources.* Nanophotonics, 2018. **7**(11): p. 1795-1817.
13. Loranger, S., P.S.J. Russell, and D. Novoa, *Sub-40 fs pulses at 1.8 µm and MHz repetition rates by chirp-assisted Raman scattering in hydrogen-filled hollow-core fiber.* Journal of the Optical Society of America B, 2020. **37**(12): p. 3550-3556.
14. Bischel, W.K. and M.J. Dyer, *Wavelength dependence of the absolute Raman gain coefficient for the Q(1) transition in H2.* Journal of the Optical Society of America B, 1986. **3**(5): p. 677-682.
15. Gladyshev, A.V., et al., *4.4-µm Raman laser based on hollow-core silica fibre.* Quantum Electronics, 2017. **47**(5): p. 491-494.
16. Benoit, A., et al. *High power Raman-converter based on H2-filled inhibited coupling HC-PCF.* in *SPIE LASE.* 2017. San Francisco: SPIE.
17. Benabid, F., et al., *Stimulated Raman Scattering in Hydrogen-Filled Hollow-Core Photonic Crystal Fiber.* Science, 2002. **298**(5592): p. 399.
18. Abdolvand, A., et al., *Solitary Pulse Generation by Backward Raman Scattering in H2-Filled Photonic Crystal Fibers.* Physical Review Letters, 2009. **103**(18): p. 183902.

19. Chen, Y.-H., J. Moses, and F. Wise, *Femtosecond long-wave-infrared generation in hydrogen-filled hollow-core fiber.* Journal of the Optical Society of America B, 2023. **40**(4): p. 796-806.
20. Couny, F., et al., *Generation and Photonic Guidance of Multi-Octave Optical-Frequency Combs.* Science, 2007. **318**(5853): p. 1118.
21. Bauerschmidt, S.T., et al., *Broadband-tunable LP01 mode frequency shifting by Raman coherence waves in a H2-filled hollow-core photonic crystal fiber.* Optica, 2015. **2**(6): p. 536-539.
22. Mridha, M.K., et al., *Thresholdless deep and vacuum ultraviolet Raman frequency conversion in hydrogen-filled photonic crystal fiber.* Optica, 2019. **6**(6): p. 731-734.
23. Tyumenev, R., et al., *Tunable and state-preserving frequency conversion of single photons in hydrogen.* Science, 2022. **376**(6593): p. 621-624.
24. Hayes, J.R., et al., *Anti-resonant hexagram hollow core fibers.* Optics Express, 2015. **23**(2): p. 1289-1299.
25. Wang, Y.Y., et al., *Low loss broadband transmission in hypocycloid-core Kagome hollow-core photonic crystal fiber.* Optics Letters, 2011. **36**(5): p. 669-671.
26. Debord, B., et al., *Ultralow transmission loss in inhibited-coupling guiding hollow fibers.* Optica, 2017. **4**(2): p. 209-217.
27. Bauerschmidt, S.T., D. Novoa, and P. St.J. Russell, *Dramatic Raman Gain Suppression in the Vicinity of the Zero Dispersion Point in a Gas-Filled Hollow-Core Photonic Crystal Fiber.* Physical Review Letters, 2015. **115**(24): p. 243901.
28. Hosseini, P., et al., *Universality of Coherent Raman Gain Suppression in Gas-Filled Broadband-Guiding Photonic Crystal Fibers.* Physical Review Applied, 2017. **7**(3): p. 034021.
29. Podzorov, A. and G. Gallot, *Low-loss polymers for terahertz applications.* Applied Optics, 2008. **47**(18): p. 3254-3257.
30. Naftaly, M., R.E. Miles, and P.J. Greenslade. *THz transmission in polymer materials — a data library*. in *2007 Joint 32nd International Conference on Infrared and Millimeter Waves and the 15th International Conference on Terahertz Electronics*. 2007.
31. Woyessa, G., et al., *Zeonex microstructured polymer optical fiber: fabrication friendly fibers for high temperature and humidity insensitive Bragg grating sensing.* Optical Materials Express, 2017. **7**(1): p. 286-295.
32. Tani, F., et al., *Effect of anti-crossings with cladding resonances on ultrafast nonlinear dynamics in gas-filled photonic crystal fibers.* Photonics Research, 2018. **6**(2): p. 84-88.